\documentclass[aps,prl,twocolumn,showpacs]{revtex4}

\usepackage{graphicx}% Include figure files
\usepackage{subfigure}
\usepackage{epsfig}
\usepackage{dcolumn}% Align table columns on decimal point
\usepackage{bm,natbib}% bold math
\usepackage{amsmath,amssymb}

\begin{document}

\newcommand{\nwc}{\newcommand}
\nwc{\beq}{\begin{equation}}
\nwc{\eeq}{\end{equation}}
\nwc{\bdm}{\begin{displaymath}}
\nwc{\edm}{\end{displaymath}}
\nwc{\bea}{\begin{eqnarray}}
\nwc{\eea}{\end{eqnarray}}
\nwc{\para}{\paragraph}
\nwc{\vs}{\vspace}
\nwc{\hs}{\hspace}
\nwc{\la}{\langle}
\nwc{\ra}{\rangle}
\nwc{\del}{\partial}
\nwc{\lw}{\linewidth}
\nwc{\nn}{\nonumber}

\nwc{\pd}[2]{\frac{\partial #1}{\partial #2}}
\nwc{\zprl}[3]{Phys.Rev. Lett. ~{\bf #1},~#2~(#3)}
\nwc{\zpre}[3]{Phys. Rev. E ~{\bf #1},~#2~(#3)}
\nwc{\zjsm}[3]{J.Stat. Mech. ~{\bf #1},~#2~(#3)}
\nwc{\zepjb}[3]{Eur. Phys. J. B ~{\bf #1},~#2~(#3)}
\nwc{\zrmp}[3]{Rev. Mod. Phys. ~{\bf #1},~#2~(#3)}
\nwc{\zepl}[3]{Europhys. Lett. ~{\bf #1},~#2~(#3)}
\nwc{\zjsp}[3]{J. Stat. Phys. ~{\bf #1},~#2~(#3)}
\nwc{\zptps}[3]{Prog. Theor. Phys. Suppl. ~{\bf #1},~#2~(#3)}
\nwc{\zpt}[3]{Physics Today ~{\bf #1},~#2~(#3)}
\nwc{\zap}[3]{Adv.Phys. ~{\bf #1},~#2~(#3)}
\nwc{\zjpcm}[3]{J.Phys. Condens. Matter ~{\bf #1},~#2~(#3)}
\nwc{\zjpa}[3]{J. Phys. A ~{\bf #1},~#2~(#3)}
\nwc{\zsci}[3]{Science ~{\bf #1},,~#2~(#3)}
\nwc{\zssc}[3]{Surf. Sci.~{\bf #1},~#2~(#3)}

\nwc{\znl}[3]{Nano Lett.~{\bf #1},~#2~(#3)}

\nwc{\zasc}[3]{Applied Surf. Sci.~{\bf #1},~#2~(#3)}
\nwc{\zjvsta}[3]{J. Vac. Sci. Technol. A ~{\bf #1},~#2~(#3)}
\nwc{\zprb}[3]{Phys. Rev. B ~{\bf #1},~#2~(#3)}
\nwc{\zjnn}[3]{J.Nanosci. Nanotechnol. ~{\bf #1},~#2~(#3)}
\nwc{\zssr}[3]{Surf.Sci.Rep. ~{\bf #1},,~#2~(#3)}
\nwc{\ztsf}[3]{Thin Solid films. ~{\bf #1},,~#2~(#3)}
\nwc{\zapl}[3]{Appl. Phys. Lett.~{\bf #1},~#2~(#3)}
\nwc{\zjcp}[3]{J.Chemical Phys.~{\bf #1},~#2~(#3)}

\nwc{\zapa}[3]{Applied Phys. A.~{\bf #1},~#2~(#3)}
\nwc{\znat}[3]{Nature(London).~{\bf #1},~#2~(#3)}

\nwc{\zjjap}[3]{J. J. Appl. Phys.~{\bf #1},~#2~(#3)}
\nwc{\zsse}[3]{Solid State Electron ~{\bf #1},~#2~(#3)}
\nwc{\zjetpl}[3]{JETP Lett.~{\bf #1},~#2~(#3)}
\nwc{\zjap}[3]{J. Appl. Phys.~{\bf #1},~#2~(#3)}
\nwc{\zntech}[3]{Nanotechnology.~{\bf #1},~#2~(#3)}
\nwc{\zjpcs}[3]{J.Phys. Conf. Series ~{\bf #1},~#2~(#3)}
\nwc{\zcsc}[3]{Curr. Sci ~{\bf #1},~#2~(#3)}

\title{Universality in Shape Evolution of Si$_{1-x}$Ge$_{x}$ Structures on High Index Silicon Surfaces}

\author{J. K. Dash, T. Bagarti, A. Rath, R. R. Juluri and P. V. Satyam\email{}}

\date{\today}

\email{pvsatyam22@gmail.com,satyam@iopb.res.in}

\affiliation{\vspace{0.5cm}Institute of Phsyics,  Sachivalaya
Marg, Bhubaneswar - 751005, India}

\begin{abstract}%\large
The MBE grown Si$_{1-x}$Ge$_x$ islands on reconstructed high index
surfaces, such as, Si(5 5 12),\\ Si(5 5 7) and Si(5 5 3) show a
universality in the shape evaluation and the growth exponent
parameters, \emph{irrespective} of the substrate orientations and
size of the island structures. This phenomena has been explained
by incorporating a deviation parameter ($\epsilon$) to the surface
barrier term ($E_D$) in the kinematic Monte Carlo (kMC)
simulations as one of the plausible mechanisms.
\end{abstract}


\maketitle{}

%\section{Introduction}
The implementation of Ge and Si$_{1-x}$Ge$_x$ nanostructures into
Si-based devices is of great potential for future high-speed
devices, due to advantages like enhanced carrier mobility and
smaller bandgap \cite{wu02,lee05}. Growth of Ge islands on clean
Si substrates (particularly on low-index-oriented) has been
extensively studied as a model system to understand hetero-epitaxy
and Stranski-Krastanov (SK) growth mechanism \cite{mo90,pcti10}.
But, the study of Ge growth on high index surfaces, such as Si (5
5 12), Si(5 5 7) and Si(5 5 3) is an area where very limited
research work has been carried out \cite{ohm05,kim08}. Previous
study by Kim \textit{et al} \cite{kim08} reported on relatively
thinner Ge growth on Si (5 5 12) and did not observe any shape
transformation of the SiGe structures. For the Ge on Si system
strain relief and diffusion play a role in determining the
morphology and composition of Ge or SiGe structures
\cite{mo90,pcti10,bzeir07}. In the present work, we have observed
the variation in the size of the Si$_{1-x}$Ge$_x$ structures
depending on the substrate orientation while having the similar
aspect ratios and growth exponent values for all three substrate
orientations. The variation in the size of the islands has been
explained by considering the experimentally observed composition
of the structures and strain associated in them. The universal
nature of shape evaluation and exponent values are simulated with
2D kMC simulations. With the interesting reconstructions on high
index surfaces and anisotropic diffusion dominating during the
direct current heating conditions, the present work stimulates the
self-assembly growth related work in the area of nanoscience and
nanotechnology.

Reconstructed high index silicon surfaces consisting of
alternating terraces and atomic steps can be used as templates to
form aligned one dimensional (1D) nanostructures
\cite{bas95,bas01,ahn02}. Atomic steps on the high index silicon
surfaces are responsible for many surface dynamic processes like
surface migration and step diffusion. It is possible to use the
high index substrates for the growth of self-organized
nanostructures \cite{ohm05}. Among the high index silicon
surfaces, Si(5 5 12) is oriented 30.5$^{\circ}$ away from (001)
towards (111) with one-dimensional periodicity over a large unit
cell width \cite{bas95,kim07}. Si(5 5 7) with vicinal angle of
9.45$^{\circ}$ from (111) towards ($11\bar{2}$ ) \cite{kosian01}
and  Si(5 5 3) , tilted at -12.27$^{\circ}$ from the (111) plane
towards the (0 0 1) plane \cite{hara08}, are important high index
Si surfaces. In all the above vicinal surfaces, the step edges are
along $\langle 1\bar{1}0 \rangle$ direction.

  It is known that strained epitaxial layers tend initially to grow as
dislocation-free islands and as they increase in size, may undergo
a shape transition \cite{tersoff93,bzeir07}. Below a critical
size, islands can have a compact symmetric shape. But at larger
sizes, they adopt a long shape, which allows better elastic
relaxation of the island's stress
\cite{tersoff93,bzeir07,Eahlesham90}. Tersoff and Tromp proposed a
model to explain the growth kinetics involving a shape transition
at a critical size by finding an
 expression for energy of dislocation-free strained
islands \cite{tersoff93}. Also, Heyn reported the kinetic Monte
Carlo (kMC) methods to study the influence of the anisotropy of
surface diffusion and of the binding energies \cite{heyn01}.
Kinetic Monte Carlo methods have been used  to study the formation
of nanoisland structures in a number of works
\cite{khor00,fichthorn02} ,however, shape transition phenomena has
hardly been studied.

%\section{Experimental Details}

The experiments discussed in the following were performed in
ultra-high vacuum(UHV), at a base pressure of $\sim 2.5 \times
10^{-10}$ mbar in a custom built molecular beam epitaxy (MBE)
system \cite{goswami03}.  Samples  of Si(5 5 12), Si(5 5 7) and
Si(5 5 3)  were prepared  from  p-type boron doped wafers (of
resistivity of 10 - 15 $\Omega$ cm). Substrates were degassed at
600$^{\circ}$C for about 12 hours followed by repeated flashing
(with direct current heating) for 30 sec. at a temperature of
1250$^{\circ}$C to remove the native oxide layer to obtain a clean
and well-reconstructed surface. The reconstruction has been
confirmed with in-situ reflection high energy electron
diffraction. The temperature was monitored with an infra-red
pyrometer calibrated with a thermocouple attached to the sample
holder. Ge was deposited to various thicknesses of 2 to 10
monolayer (ML) at a typical deposition rate of 0.6 ML/min at
substrate temperature 600$^{\circ}$C [with direct current
heating(DH)]. The samples were then post annealed at a temperature
of 600$^{\circ}$C for 15 minutes by DH method. Also a set of
samples were prepared with radiative heating (RH)(where heating is
achieved through a filament underneath).The post growth
characterization of the samples was carried out ex-situ by field
emission gun based scanning electron microscopy (FEGSEM) with 20
kV electrons.

\begin{figure}
\centering \vspace{1cm} \epsfig{file=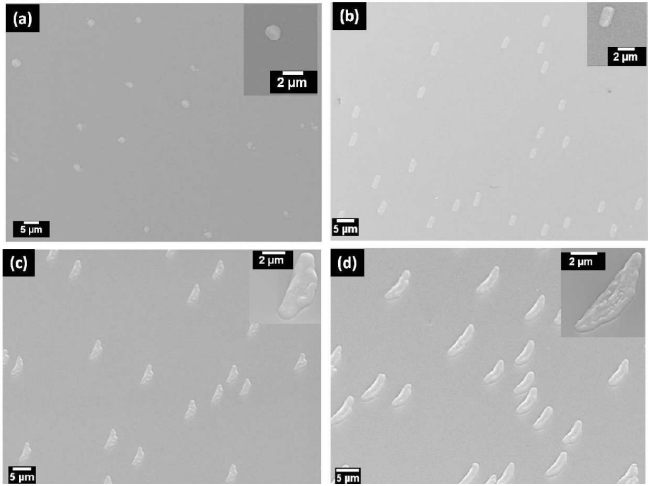, width=\lw}
\caption{ SEM micrographs of (a)2 ML(b)5 ML(c)8 ML (d)10 ML
Ge/Si(5 5 12) at 600$^{\circ}$C-DH. Insets show magnified image of
one structure.} \label{fig1}
\end{figure}

\begin{figure}
\centering \vspace{1cm} \epsfig{file=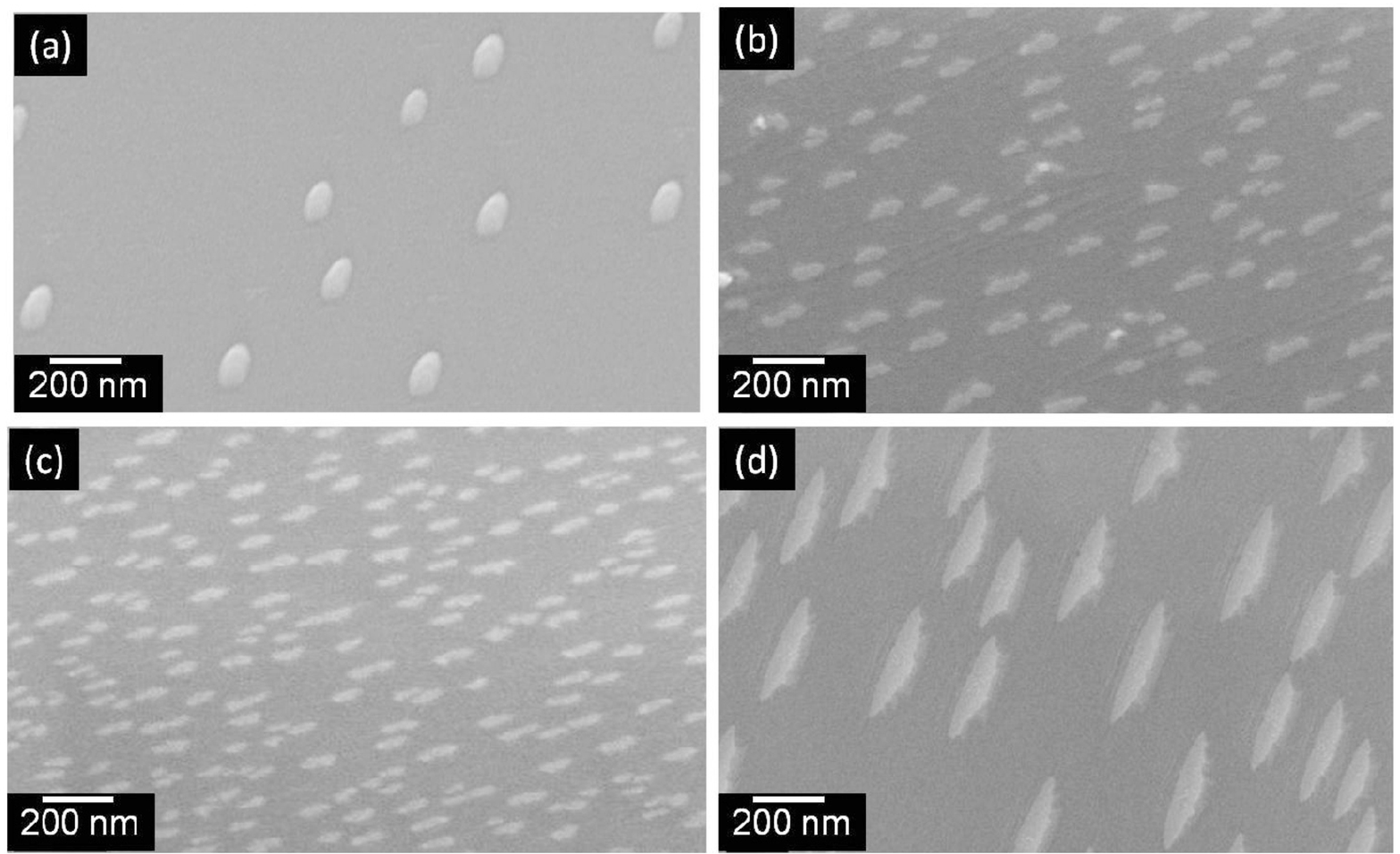, width=\lw}
\caption{ SEM micrographs of (a)3 ML(b)5 ML(c)8 ML (d)10 ML
Ge/Si(5 5 7) at 600$^{\circ}$C-DH} \label{fig2}
\end{figure}

\begin{figure}
\centering \epsfig{file=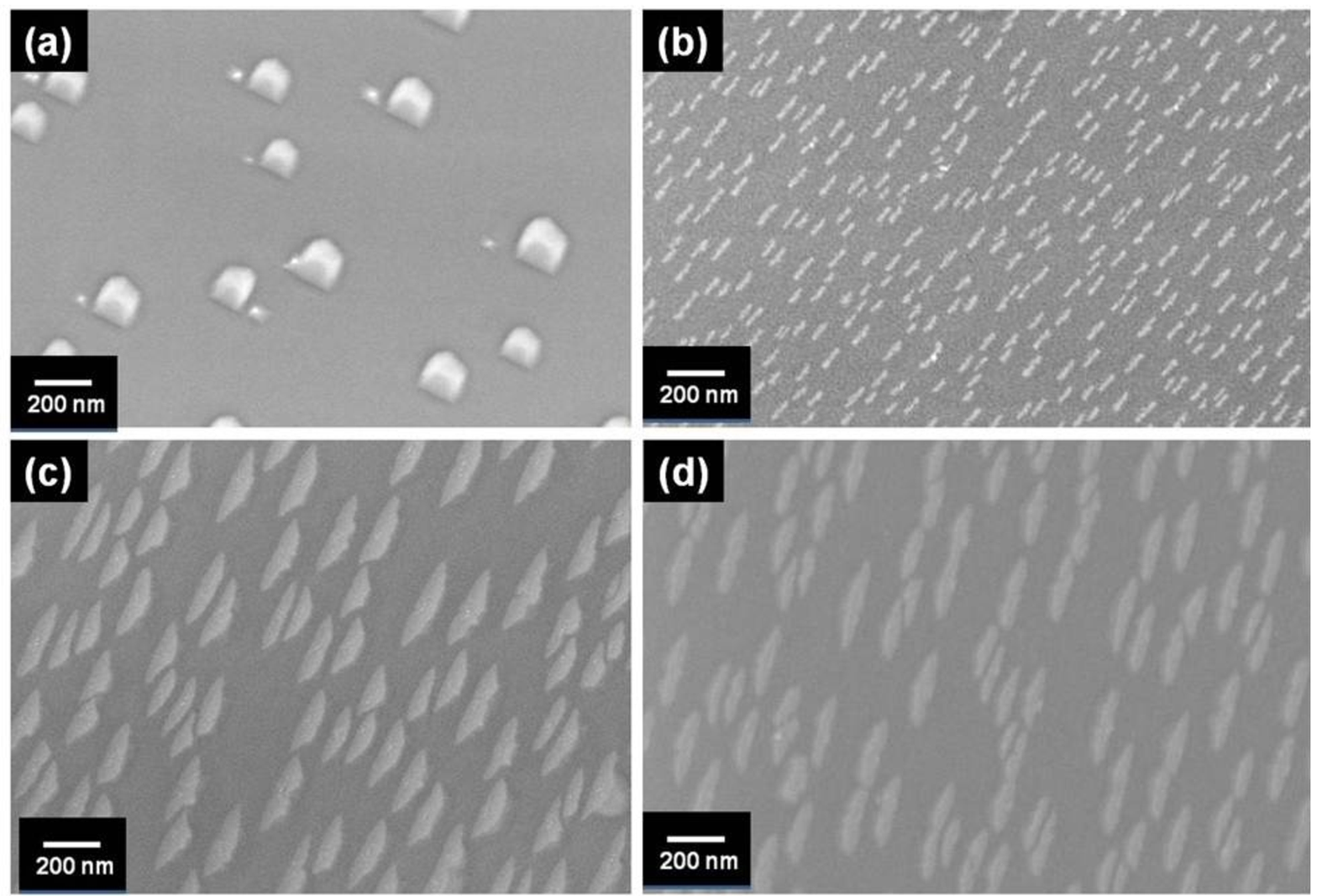, width=\lw} \caption{SEM
micrographs of (a)3 ML(b)5 ML(c)8 ML (d)10 ML Ge/Si(5 5 3) at
600$^{\circ}$C-DH} \label{fig3}
\end{figure}

\begin{figure}
\centering \epsfig{file=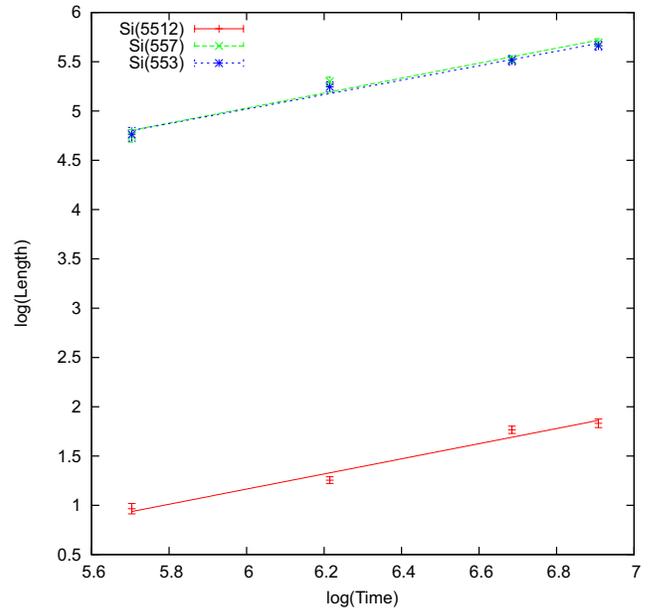, width=\lw} \caption{Plot showing
the Length Exponents of the strucrures formed on Si(5 5 12), Si(5
5 7)and Si(5 5 3) surfaces }\label{fig4}
\end{figure}

\begin{table}
\caption{\label{tab:table1}Mean Length (L)}
\begin{ruledtabular}
\begin{tabular}{cccc}
 Thickness&Ge/Si(5 5 12) &Ge/Si(5 5 7)&Ge/Si(5 5 3)\\
(ML)&Mean L in ($\mu$m)& Mean L in(nm)& Mean L in(nm)\\\hline
 3&2.63 $\pm$ 0.14  &115.5 $\pm$ 7.2&117.5 $\pm$ 8.2 \\
 5&3.51 $\pm$ 0.12  &200.6 $\pm$ 8.7&189.8 $\pm$ 9.7\\
 8&5.85 $\pm$ 0.22  &250.6 $\pm$ 9.6&248.7 $\pm$ 11.6\\
 10&6.25 $\pm$ 0.27  &297.8 $\pm$11.4&288.3 $\pm$ 12.4\\

\end{tabular}
\end{ruledtabular}
\end{table}

\begin{table}
\caption{\label{tab:table2} Aspect Ratios (length/width)}
\begin{ruledtabular}
\begin{tabular}{cccc}
 Thickness&Ge/Si(5 5 12) &Ge/Si(5 5 7)&Ge/Si(5 5 3)\\
(ML)&Aspect Ratio&Aspect Ratio&Aspect Ratio\\\hline
 3&2.10 $\pm$ 0.1&1.79 $\pm$ 0.2&1.74 $\pm$ 0.6\\
 5&2.20 $\pm$ 0.1&2.28 $\pm$ 0.6&2.34 $\pm$ 0.8\\
 8&2.84 $\pm$ 0.5&2.88 $\pm$ 0.3&2.92 $\pm$ 0.9\\
 10&3.13 $\pm$ 0.3&3.17 $\pm$ 0.4&3.15 $\pm$ 0.5\\

\end{tabular}
\end{ruledtabular}
\end{table}

\begin{table}
\caption{\label{tab:table3}Length Exponent ($\alpha$)}
\begin{ruledtabular}
\begin{tabular}{ccccc}
 Exponent &Ge/Si(5 5 12) &Ge/Si(5 5 7)&Ge/Si(5 5 3)&from kMC\\\hline
 DH&0.77$\pm$0.04&0.76$\pm$0.03&0.73$\pm$0.02&0.60$\pm$0.02 \\
RH&0.51$\pm$0.03&0.53$\pm$0.02&0.52$\pm$0.04&0.45$\pm$0.02 \\

\end{tabular}
\end{ruledtabular}
\end{table}

Figure \ref{fig1} (a) and (b) depict the FEGSEM image for 2 ML and
5ML thick Ge film deposited on Si(5 5 12) substrate. As shown in
Fig \ref{fig1}(a), the island structures are found to be spherical
shape for a 2 ML Ge deposition, but for 5 ML Ge growth,
rectangular island structures have been observed. For the 8ML and
10ML deposition cases,  we observe trapezoidal structures
[Fig.\ref{fig1}(c),(d)]. It was reported earlier that DH is a
necessary condition to form trapezoidal structural structures on
Si(5 5 12) \cite{dash11}. The composition of these nanostructures
has been characterized by using STEM - EDS
\cite{dash11,dashmsm11}, Rutherford backscattering spectrometry
(RBS)\cite{dash11} and the synchrotron-based high resolution x-ray
diffraction (HRXRD). The HRXRD showed presence of graded
Si$_{1-x}$Ge$_x$ system for the Si(5 5 12)and is also confirmed by
RBS measurements. While RBS measurements show no prominent graded
Si$_{1-x}$Ge$_x$ structures for the case of Si(5 5 3)and Si(5 5 7)
substrate orientations.

In fig.\ref{fig2}, Ge growth of overlayer 3ML to 10ML on Si(5 5 7)
has been shown where spherical nanoislands and rectangular nano
rods are formed for 3ML and 5ML thicknesses[Fig.\ref{fig2}(a),(b)]
and nano trapezoid in the case of 8 ML and 10 ML Ge growth
[Fig.\ref{fig2}(c), (d)]. We have maintained the same growth
condition for Si (5 5 3) substrate and seen that the growth
kinetics and shape transformation follow in a similar way
resulting from rectangular nano rod to nano trapezoid, as shown in
fig.\ref{fig3}.

We have measured the length and aspect ratio (length/width) of all
the Ge-Si structures which has been shown in a tabular form in
table \ref{tab:table1} and table \ref{tab:table2}. In the case of
Ge/Si (5 5 12) system, the size of the aligned structures are of
micrometer size. But for the Si(5 5 7) and Si(5 5 3) systems,the
size of the aligned structures are of nanometer size, though the
shape evolution and also the aspect ratios  are similar to that of
Si(5 5 12) . From table \ref{tab:table1}, it is clearly seen that
with increasing thickness, the size and aspect ratio of the Ge-Si
structures increase accordingly. The aspect ratios of  the Si-Ge
elongated structures on Si( 5 5 12),Si(5 5 7) and Si(5 5 3)
increase in a similar fashion as a function of increasing Ge
growth coverages, which has been shown in table \ref{tab:table2}.

The growth of island followed a universality and we found from our
experiments that the mean length of islands $\langle L \rangle
\sim t^\alpha$ where $t$ is the time of deposition. The growth
exponent from the experimental data is found to be about 0.75
$\pm$ 0.02 for the case of the direct heating and  0.5 $\pm$ 0.01
for the radiative heating case in all the three systems [table
\ref{tab:table3}].We observe that since $\alpha < 1 $, the growth
of island is sublinear. In fig \ref{fig4}, the log-log plot for
size(length) vs deposition time shows almost same slope($\alpha$)
irrespective of the size of the Si-Ge structures on the three
substrates. We now disscuss the shape transformation of structures
by kMC.

%\section{Theoretical Modeling}
 The kinetic Monte Carlo simulations were performed on a $L\times
L$ square lattice with $L=100, 200, 300, 400$. A coordinate system
has been chosen such that the x-axis is directed along
perpendicular to the step edges. We define horizontal bond and
vertical bond as a pair of nearest neighbor atoms having the same
y coordinates and same x coordinates respectively. Similarly a
vertical bond is defined by a nearest neighbor pair of atoms with
same x coordinates. The hopping rate of an adatom is given by
\begin{equation}
w = \nu_0 e^{- (E_D + n_1 E_1 + n_2 E_2 + n_3 E_3) / k_B T}
\end{equation}
where $k_B$ is the Boltzman constant, $T$ is the temperature and
$\nu_0 = 2k_BT/h$ is the vibrational frequency in the direction of
the hopping. A typical value for our case is $\nu_0 =
3.6\times10^{13}$ sec$^{-1}$. The quantities $E_1$, $E_2$ and
$E_3$ denote the binding energies for the horizontal bond,
vertical bond and the next nearest neighbor interaction
respectively ; $n_1$,$n_2$ and $n_3$ are the number of horizontal
bonds, vertical bonds and  the number of next nearest neighbors
respectively and $E_D$ is a surface barrier term. For $E_D$ and
$E_i$, ($i=1,2,3$) constant, we obtained spherically symmetric
island structures in the kMC simulation, since the binding
energies are same for all bonds  and surface barrier term is
independent of direction. To obtain asymmetrical structures , we
need to break the hopping symmetry. In our model, we introduce
anisotropy through binding energies of different types of bonds
and the dependence of surface barrier on the direction of hopping.
For the isotropic cases, however, the hopping rate depends only on
the nearest and next nearest neighbor interactions but does not
depend on the specific arrangement of the neighboring atoms. In
our experiment, the shape transformation has been observed when Ge
is grown under DH condition \cite{dash11}. To simulate the
experimentally observed shape variations, the surface barrier term
is modified, so that it allows an asymmetric hopping of an adatom
along the direction perpendicular to the step edge ( i.e. $\pm$ x
direction). The values of $E_D$ along the step edges are same in
both $\pm$ y directions. When $E_1 = E_2$ and $E_D$ does not
depend on the direction of hopping,spherical islands are formed.
Symmetric elongated structures are formed for $E_2 > E_1$ and
$E_D$ uniform in all the directions of hopping. Trapezoidal
islands can only be obtained when an asymmetric hopping term is
present. The values of $E_D$ are defined as follows.a) $E_D =
E_0(1+\epsilon)$ for hopping along the positive x-direction, b)
$E_D = E_0(1-\epsilon)$ for hopping along the negative
x-direction, c) $E_D = E_0$ for hopping along y-axis where
$\epsilon$ depends only on the magnitude of the current (but for
the simulations, this is a parameter only). Also $0 < \epsilon <
1$ so that the probabilities are strictly less than unity. Note
that when $\epsilon =0$, it reduces to the case of anisotropic
island formation for $E_2 > E_1$ as in the case of
Ref.\cite{heyn01} and isotropic islands formation for $E_2 = E_1$
as in Ref. \cite{lo99}.Therefore, in our kMC simulation a nonzero
value of $\epsilon$ should correspond to the MBE growth done under
DC heating. The mean shape of islands for this case is found to be
trapezoids.

We follow the kMC algorithm in \cite{lo99,lo991}. We set the total
coverage $\theta$ and the total number of Monte carlo steps
$N_{\mbox{\scriptsize MC}}$. Particles are deposited at a constant
flux $\theta/N_{\mbox{\scriptsize MC}}$. We define a dimensionless
scale parameter $\phi=E_0/K_BT$, where $E_0$ is the surface energy
barrier of the system when $\epsilon=0$.  This sets the time step
for the simulation. An atom with empty adjacent site is called
'active'.
 An active atom is chosen at random and a single diffusion event is allowed to occur with a probability consistent with Eq.(1).
 If an atom is already there on the site to which it chooses to hop, then hopping fails.
 Time is incremented irrespective of whether the hopping is successful or not.
 The following parameters were used for our kMC : $\phi=1.0, E_1=1.0, E_2=6.0 ,
E_3 =0.5, E_0=1.0, \epsilon = 0.8$, and coverage is 2.5$\%$. It is
important to note that the relative strength of the energy values
is crucial in determining the shape of the structures. The above
set of energy parameters is one such example.

\begin{figure}
\centering \vspace{0cm} \epsfig{file=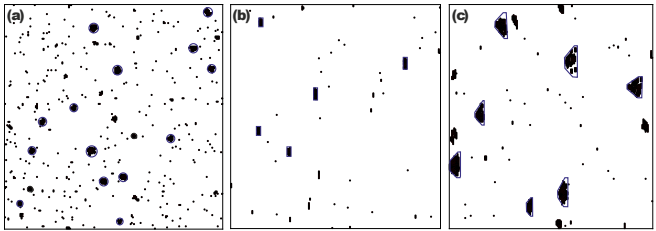, width=\lw}
\caption{ Snapshot of kMC island growth (a)spherical shaped
islands at $\epsilon = 0$, $E_1 = E_2$ (b) $\epsilon= 0.8$ ,$ E_1
< E_2$ rod like structures at coverage $0.2 \theta$ (c) $\epsilon
= 0.8$ , $E_1 < E_2$ showing shape transition by forming
trapezoidal island structures at coverage $\theta$.} \label{fig5}
\end{figure}

 In fig. \ref{fig5} (a),spherical shaped islands for $\epsilon = 0, E_1 = E_2$
are seen. For $\epsilon = 0$, $E_1 < E_2$ at coverage $0.2
\theta$, rod like aligned structures formed [fig.\ref{fig5}(b)].
Figure\ref{fig5} (c) is final snapshot of the trapezoidal
structures formed at coverage $\theta$ for $\epsilon = 0.8$ and
$E_1 < E_2$. From the kMC simulation, we have found $\alpha = 0.6
\pm 0.02 $(for $\epsilon=0$) and  $\alpha = 0.45 \pm 0.03 $(for
$\epsilon=0.8$).

%\section{Conclusion}
In summary,we report the shape evolution of MBE grown Si$_{1-
x}$Ge$_x$ islands on reconstructed high index Si(5 5 12),   Si(5 5
7) and Si(5 5 3) surfaces. We show that a self assembled growth at
optimum thickness leads to interesting shape transformations,
namely, spherical islands to rectangular nanostructures and then
to elongated trapezoidal structures. We have experimentally
observed an universality in the growth of the islands for all
three high index surfaces by evaluating the aspect ratios and
growth exponent.The growth exponent (for the longer side of the
structures) experimentally found to be about 0.75 $\pm$ 0.02 for
the case of the direct heating and 0.5 $\pm$ 0.01 for the
radiative heating case in all the three systems. Our kMC
simulations show that such variations can be understood by
introducing a  deviation parameter $\epsilon$ in surface barrier
term $E_D$. The experimentally observed shape variations and the
growth exponent values are in good agreement with kMC simulations.
This suggests the role of stochastic process involved in the shape
transitions of nanoscale structures.
%\section{Acknowledgement}
We thank S. D.Mohanty for stimulating discussions. PVS would like
to thank the Department of Atomic Energy, Government of India for
granting FEGSEM under 11th plan.

\end{document}